\documentclass[a4paper,10pt,twiside]{zhepnp}
\usepackage{multicol}
\usepackage{graphicx}
\usepackage{booktabs}
\usepackage{amssymb,bm,mathrsfs,bbm,amscd}
\usepackage[tbtags]{amsmath}
\usepackage{lastpage}
\usepackage{textcomp}
\usepackage{hyperref}
\usepackage{xcolor}
\usepackage{CJK}
\usepackage{authblk}

\newcommand \ssr {Space Sci. Rev.}
\newcommand \aap { Astron. Astrophys.}
\newcommand \apj {Astrophys. J.}
\newcommand \apjl {Astrophys. J. Lett.}
\newcommand \apjs {Astrophys. J. Suppl.}
\newcommand \solphys {Sol. Phys.}
\newcommand \lrsp {Living Rev. Sol. Phys.}

\newcommand \raa {Research in Astronomy and Astrophysics}
\newcommand \mnras {Monthly Notices of the Royal Astronomical Society}

\begin{document}

\title{Simultaneous detection of flare-associated kink oscillations and extreme-ultraviolet waves\thanks{lidong@pmo.ac.cn}}

\author{Dong~Li$^{1,2}$,Zhenyong~Hou$^3$, Xianyong~Bai$^{4,5}$, Chuan~Li$^{6,7}$, Matthew~Fang$^{8}$, Haisheng~Zhao$^9$, Jincheng~Wang$^{2,10}$, and Zongjun~Ning$^1$}
 \affil{$^{1}$Key Laboratory of Dark Matter and Space Astronomy, Purple Mountain Observatory, CAS, Nanjing 210023, China}
 \affil{$^{2}$Yunnan Key Laboratory of the Solar physics and Space Science, Kunming 650216, China}
 \affil{$^{3}$School of Earth and Space Sciences, Peking University, Beijing, 100871, China}
 \affil{$^{4}$National Astronomical Observatories, Chinese Academy of Sciences, Beijing 100101, China}
 \affil{$^{5}$School of Astronomy and Space Sciences, University of Chinese Academy of Sciences, Beijing 100049, China}
 \affil{$^{6}$School of Astronomy and Space Science, Nanjing University, Nanjing 210023, China}
 \affil{$^{7}$Key Laboratory of Modern Astronomy and Astrophysics (Nanjing University), Ministry of Education, Nanjing 210023, China}
 \affil{$^{8}$Columbus Academy, Gahanna Ohio, OH 43230, USA}
 \affil{$^{9}$Key Laboratory of Particle Astrophysics, Institute of High Energy Physics, CAS, Beijing 100049, China}
 \affil{$^{10}$Yunnan Observatories, Chinese Academy of Sciences, Kunming Yunnan 650216, China}
\maketitle

\begin{abstract}
Kink oscillations, which are frequently observed in coronal loops
and prominences, are often accompanied by extreme-ultraviolet (EUV)
waves. However, much more needs to be explored regarding the causal
relationships between kink oscillations and EUV waves. In this
article, we report the simultaneous detection of kink oscillations
and EUV waves that are both associated with an X2.1 flare on 2023
March 03 (SOL2023-03-03T17:39). The kink oscillations, which are
almost perpendicular to the axes of loop-like structures, are
observed in three coronal loops and one prominence. One short loop
shows in-phase oscillation within the same period of 5.2~minutes at
three positions. This oscillation could be triggered by the pushing
of an expanding loop and interpreted as the standing kink wave. Time
lags are found between the kink oscillations of the short loop and
two long loops, suggesting that the kink wave travels in different
loops. The kink oscillations of one long loop and the prominence are
possibly driven by the disturbance of the CME, and that of another
long loop might be attributed to the interaction of the EUV wave.
The onset time of the kink oscillation of the short loop is nearly
same as the beginning of an EUV wave. This fact demonstrates that
they are almost simultaneous. The EUV wave is most likely excited by
the expanding loop structure and shows two components. The leading
component is a fast coronal wave, and the trailing one could be due
to the stretching magnetic field lines.
\end{abstract}


\section{Introduction}
Kink oscillations, which are commonly observed as transverse
oscillations of loop-like systems, are often perpendicular to the
loop axis, non-axisymmetric, and hardly compressive in the
long-wavelength regime. They are often associated with
magnetohydrodynamic (MHD) waves in the solar atmosphere at multiple
heights, and thus are believed to play a key role in coronal heating
\cite{Van20,Nakariakov21,Chelpanov22}. The kink-mode oscillation was
first discovered as the cyclic transverse motions of coronal loop,
which was shown as a large-amplitude oscillation (i.e., $\gg$1~Mm)
with rapid decay, regarded as `decaying oscillation'
\cite{Aschwanden99,Nakariakov99}. The decaying kink oscillation only
persists over several wave periods
\cite{Su18,Nechaeva19,Kumar22,Li23a}, and is always associated with
an impulsive eruption, for instance, the decaying kink oscillation
could be induced by a solar flare, a coronal jet, an
extreme-ultraviolet (EUV) wave, and so on
\cite{Zimovets15,Reeves20,Zhang22}. The decaying time $\tau$ is
roughly equal to several oscillation periods ($P$), for instance, it
is found that the average damping time is $\tau=1.79\times P$ for
large-amplitude decaying oscillations
\cite{Nechaeva19,Nakariakov23}. On the other hand, the decayless
kink oscillations, which are identified as small-amplitude ($<$1~Mm)
oscillations without significant decay, are also observed as the
transverse displacement oscillations of loop-like structures
\cite{Anfinogentov15,Li20,Mandal21,Shi22,Zhong23}. Observations
suggest that the decayless kink oscillations are persistent in solar
multi-height atmospheres, for instance, they are frequently detected
in coronal loops \cite{Safna22,Li23}, solar prominences
\cite{Arregui18,Li18a}, high-temperature flare loops
\cite{Li18b,Shi23}, coronal bright points \cite{Gao22}, and the
sunspot \cite{Yuan23}, which could be related to their magnetic
nature. The kink-mode oscillations, including decaying and decayless
oscillations, are usually observed as the spatial displacement
oscillations in image sequences \cite{Goddard16,Li22a,Li22b,Guo22},
or they are detected as the Doppler shift oscillations in spectral
lines \cite{Tian12,Li17}. For the standing kink oscillations, their
oscillation periods and loop lengths reveal a linear increasing
relationship, and the oscillation periods could be measured from
tens of seconds to dozens of minutes
\cite{Anfinogentov15,Lid22,Ning22,Zhong22,Zhong23S,Gao23,Petrova23}.
At the same time, multiple harmonics of kink oscillations are
detected in coronal loops, namely, the fundamental and second or
third harmonics \cite{Duckenfield19,Zhang23}, which are helpful to
diagnose the density stratification of oscillating loops
\cite{Andries05}.

EUV waves, which are always observed as large-scale, fast
propagating disturbances in the solar corona, are often
characterized by the diffuse wavefronts with arc-shaped or circular
profiles \cite{Shen22}. They are commonly associated with energetic
eruptions such as solar flares, coronal jets, Type II radio bursts,
and coronal mass ejections (CMEs), namely, the wave centers always
lie closely to the epicenter of the associated eruptions
\cite{Asai12,Liu14,Warmuth15,Zheng19,Shen21,Hou23}. The EUV waves
are spectacular features in the solar corona, since they are coronal
disturbances that can expand quickly across most fraction of the
solar surface in tens of minutes, i.e., the detected propagation
speeds are in the range of $\sim$200$-$1500~km~s$^{-1}$ with an
average value of 650~km~s$^{-1}$ \cite{Nitta13,Shen22,Liu23}. In the
Atmospheric Imaging Assembly (AIA) era \cite{Lemen12}, the study of
EUV waves becomes a topic of particular interest due to their
high-resolution observations \cite{Chen11,Shen12a,Shen12b}. The
bright wavefronts can be clearly seen in passbands of AIA~211~{\AA}
and 193~{\AA}, while the dark wavefronts may appear in the passband
of AIA~171~{\AA} \cite{Warmuth15}. At the same time, two components
are found in the EUV wave, namely, one is the leading fast
magnetosonic wave, and the other is some trailing wave-like
signature \cite{Chen11,Shen14a,Sun22}. On the other hand, the
quasi-periodic, fast-mode propagating (QFP) wave trains are also
investigated intensively in the AIA era, a typical QFP event
consists of multiple coherent and concentric wavefronts, and thus
shows the quasi periodicity of a few minutes to tens of minutes
\cite{Wang22,Zhou22}. Obviously, the QFP is distinguished from an
EUV wave, and thus it is out scope of this study.

The kink oscillation is one of the most studied MHD waves, mainly
because it is easily observed on the Sun and plays a crucial role in
diagnosing plasma parameters and inferring magnetic fields in the
solar atmosphere, termed as `MHD coronal seismology'
\cite{Yuan16a,Yuan16b,Yuan16c,Yang20a,Yang20b,Anfinogentov22,Kolotkov23}.
The EUV wave has also been extensively studied, largely because it
could also be applied in the coronal seismology for diagnosing the
coronal parameters, and provides a new tool for probing the coronal
heating and particle acceleration \cite{Liu14,Shen22,Zheng23}. Kink
oscillations are usually associated with EUV waves, for instance,
the transverse oscillations of coronal loops and prominences are
frequently observed to be triggered by EUV waves
\cite{Asai12,Liu13,Shen14b,Zhang18,Devi22}. Those EUV waves always
appear after solar flares, and then cause transverse oscillations in
coronal loops or prominences. As an example, the kink oscillations
are simultaneously detected in a coronal loop and a filament, both
of which follow after a circular-ribbon flare \cite{Zhang20z}. Some
authors \cite{Shen18} report an EUV wave and a kink oscillation that
are simultaneously triggered by the course of jet-loop interaction.
However, the simultaneous detection of kink oscillations and an EUV
wave accompanied by a solar flare is rarely reported. In this paper,
we investigate the driving mechanisms of kink oscillations in three
coronal loops and one prominence, and a simultaneous EUV wave. The
paper is organized as follows: Section~2 describes the observations,
Section~3 shows our main results, Section~4 presents some
discussions, and Section~5 offers a brief summary.

\section{Observations}
We analyze kink oscillations of three coronal loops and one
prominence, and they are both associated with a solar flare occurred
on 2023 March 03. Those solar activities were located in the active
region of NOAA~13234 near the solar west limb, i.e., N21W76. They
were simultaneously measured by the Atmospheric Imaging Assembly
(AIA) \cite{Lemen12} on board the Solar Dynamics Observatory (SDO),
the Solar Upper Transition Region Imager
\href{https://sun10.bao.ac.cn/SUTRI/}{(SUTRI)} \cite{Bai23} on board
the first spacecraft of the Space Advanced Technology demonstration
satellite series (SATech-01), the Chinese H$\alpha$ Solar Explorer
\href{https://ssdc.nju.edu.cn}{(CHASE)} \cite{LiC22,LiC23}, the
Geostationary Operational Environmental Satellite (GOES), and the
Hard X-ray Modulation Telescope (Insight-HXMT) \cite{Zhangs20}, as
listed in Table~\ref{tab1}.

\begin{table*}[htbp]
\caption{Observational instruments used in this article.}
\label{tab1} \centering \tabcolsep 5pt
\begin{tabular}{c c c c c c c c}
\toprule[1.1pt]
Instruments   &  Windows    &   Wavelengths      &     Cadence      &  Pixel scale   &   Observational time   \\
\midrule[0.8pt]
              &   EUV      &     171~{\AA}       &                  &                &                         \\
SDO/AIA       &   EUV      &     193~{\AA}       &       24~s       & 0.6$^{\prime\prime}$  &   17:30$-$18:59~UT     \\
              &   EUV      &     211~{\AA}       &                  &                &                         \\
\midrule[0.8pt]
SUTRI         &  EUV       &     465~{\AA}       &     $\sim$30~s   &1.23$^{\prime\prime}$ &   17:45$-$18:40~UT    \\
\midrule[0.8pt]
              &  H$\alpha$ &  6562.8~{\AA}        &                 &                     &         \\
CHASE         &  H$\alpha$ &  6564.9~{\AA}        &   $\sim$71~s    & 1.04$^{\prime\prime}$ &   17:24$-$17:52~UT     \\
\midrule[0.8pt]
              &   SXR      &    1$-$8~{\AA}        &                &       --       &                         \\
GOES          &   SXR      &    0.5$-$4~{\AA}      &        1~s     &       --       &   17:39$-$18:20~UT     \\
\midrule[0.8pt]
HXMT/HE       &   HXR      &   100$-$600~keV       &        0.5~s   &        --      &   17:48$-$17:56~UT     \\
\bottomrule[1.1pt]
\end{tabular}
\end{table*}

SUTRI is designed to capture the full-Sun image at the spectral line
of Ne~VII~465~{\AA}, which has a formation temperature of about
0.5~MK \cite{Tian17}. Each pixel scale corresponds to
$\sim$1.23$^{\prime\prime}$, and the time cadence is about 30~s.
SDO/AIA provides full-disk solar maps at multiple EUV/UV channels
nearly simultaneously. In this study, We use the AIA maps in three
passbands of 171~{\AA}, 193~{\AA}, and 211~{\AA}, corresponding to
the temperatures of roughly 0.6~MK, 1.6~MK, and 2.0~MK
\cite{Lemen12}, respectively. In order to avoid the saturation map,
the time cadence of each AIA passband is selected as 24~s
\cite{Li15}. Those AIA maps have been calibrated by aia\_prep.pro,
and thus have a spatial scale of 0.6$^{\prime\prime}$~pixel$^{-1}$.
CHASE acquires the spectroscopic observation in wavebands of
H$\alpha$ and Fe~I. In this study, we use the full-Sun spectral
images at 6562.8~{\AA} and 6564.9~{\AA}, corresponding to the line
core and wing of H$\alpha$. They have a pixel scale of
$\sim$1.04$^{\prime\prime}$ and a time cadence of $\sim$71~s. We
also use the X-ray light curve recorded by GOES and the High Energy
X-ray telescope (HE) on board Insight-HXMT, which have time cadences
of 1.0~s and 0.5~s, respectively.

\section{Results}
\subsection{Overview}
\begin{figure}
\centering
\includegraphics[width=0.8\linewidth,clip=]{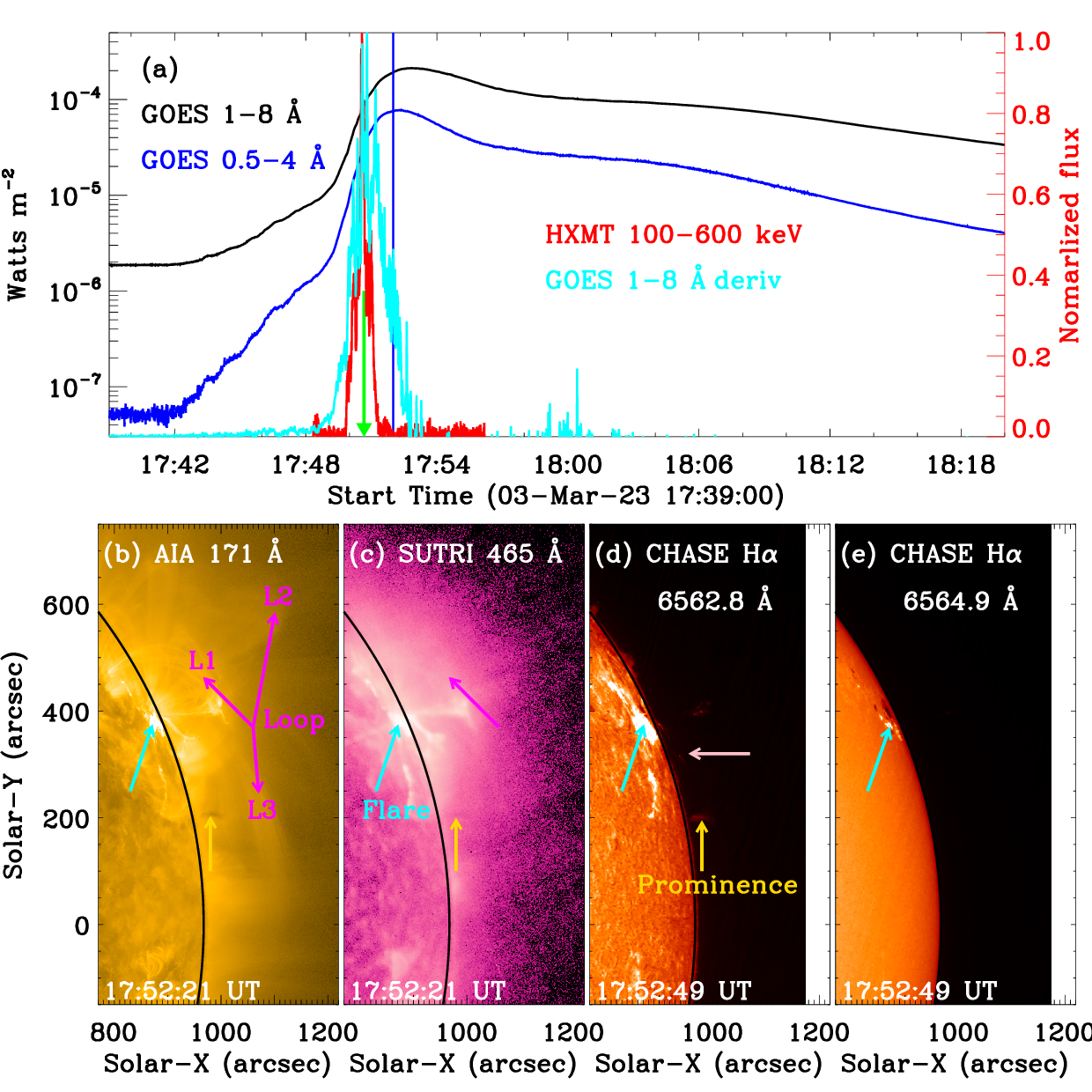}
\caption{Overview of the eruptive events on 2023 March 03. (a):
X-ray light curves between 17:39~UT and 18:20~UT at channels of
GOES~1$-$8~{\AA} (black), its derivative (cyan) and 0.5$-$4~{\AA}
(blue), and HXMT~100$-$600~keV (red). The vertical blue line marks
the SXR peak time of the X2.1 flare, and the green arrow indicates
the HXR peak time. (b)$-$(e): Multi-wavelength snapshots with a FOV
of $\sim$450$^{\prime\prime}$$\times$900$^{\prime\prime}$ measured
by SDO/AIA at 171~{\AA} (b), SUTRI at 465~{\AA} (b), CHASE/HIS at
H$\alpha$~6562.8~{\AA} (d) and 6564.9~{\AA} (e), respectively. The
color arrows indicate the targeted coronal loops (magenta), the
solar flare (cyan), the prominence (gold), and a longer but weaker
loop (pink), respectively. \label{over}}
\end{figure}

Figure~\ref{over}~presents the overview of the targeted coronal
loops and the associated solar flare and prominence on 2023 March 3.
Panel~(a) shows SXR light curves recorded by GOES at 1$-$8~{\AA}
(black) and 0.5$-$4.0~{\AA} (blue) from 17:39~UT to 18:20~UT,
indicating an X2.1-class flare. It started at $\sim$17:39~UT and
peaked at about 17:52~UT
\href{https://www.solarmonitor.org/?date=20230303}{(SolarMonitor)},
as marked by the vertical blue line. We notice that the HXR flux
recorded by HXMT/HE (red) agrees well with the SXR derivative flux
at GOES~1$-$8~{\AA} (cyan), suggesting the Neupert effect
\cite{Neupert68,Ning08,Ning09} in the X2.1 flare.

Figure~\ref{over}~(b) and (c) show EUV maps with a field-of-view
(FOV) of about 450$^{\prime\prime}$$\times$900$^{\prime\prime}$ in
passbands of AIA~171~{\AA} and SUTRI~465~{\AA} at about the SXR peak
time of the X2.1 flare (i.e., $\sim$17:52~UT), panels~(d) and (e)
present the H$\alpha$ maps with the same FOV taken by CHASE in
wavelengths of 6562.8~{\AA} and 6564.9~{\AA}. A series of coronal
loops can be seen at AIA~171~{\AA}, and three groups of them are
used to study the kink oscillations, as marked by the magenta
arrows. We notice that each group appears to consist of multiple
blended loops, which is impossible to be distinguished. So, they are
regarded as loop systems, and their fine-scale structures are not
considered. The selected loop systems seem to reveal different
features, i.e., a short compact loop (L1), and two long diffuse
loops (L2 and L3). It seems that those three loops root in the same
region, namely, their footpoints are located closely to the X2.1
flare, as indicated by the cyan arrow. The X2.1 flare can be clearly
observed in those four maps, but the coronal loops can be only seen
in EUV maps. In particular, the coronal loops of L2 and L3 are not
detected by SUTRI, mainly because that they are diffuse and weak. A
small solar prominence can be seen underlying the long loop L3, and
it shows a dark feature in passbands of AIA~171~{\AA} and
SUTRI~465~{\AA}, but it is bright in the H$\alpha$ map observed by
CHASE at 6562.8~{\AA}, which is consistent with the low-temperature
nature of the prominence \cite{Parenti14}. Moreover, the solar
prominence is likely the part of a longer loop connecting the
eruption center, as indicated by the pink arrow in
Figure~\ref{over}~(d).

\subsection{Kink oscillations}
\begin{figure}
\centering
\includegraphics[width=0.8\linewidth,clip=]{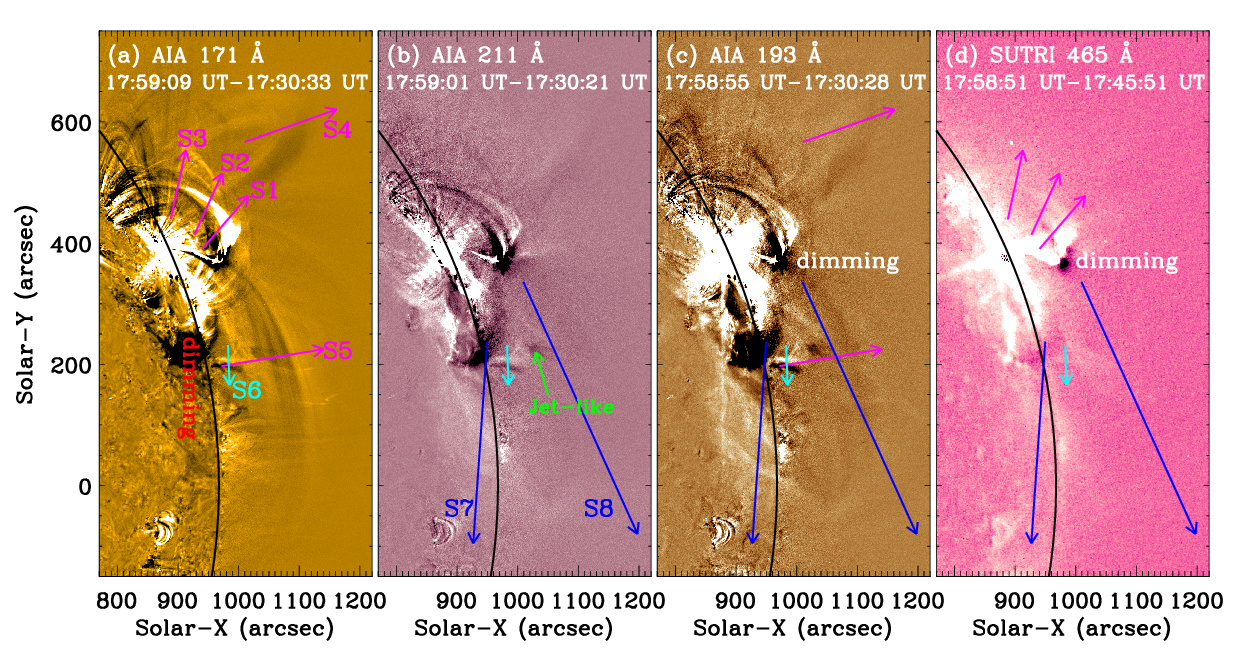}
\caption{Base-diffidence maps in passbands of AIA~171~{\AA} (a),
211~{\AA} (b) and 193~{\AA} (c), and SUTRI~465~{\AA} (d). The color
arrows mark the locations of eight artificial slits (S1$-$S8), which
are used to generate time-distance maps. The green arrow indicates a
jet-like structure. Two animations that show the evolution of the
transverse oscillations and EUV waves are available, which cover a
duration of $\sim$27~minutes from 17:48~UT to 18:15~UT. \label{img}}
\end{figure}

In Figure~\ref{over}~(b) and (c), we can see that those coronal
loops are very fuzzy, mainly because the diffuse nature of EUV
emissions. In order to highlight those loops, the base-diffidence
maps are derived from the origin maps, as shown in Figure~\ref{img}.
The coronal dimming can be simultaneously found in passbands of
SUTRI~465~{\AA}, AIA~171~{\AA}, 211~{\AA}, and 193~{\AA}, and the
short loop (L1) can be seen in those four maps. Two long diffuse
loops can be clearly seen in the passband of AIA~171~{\AA}, and they
can also be seen at AIA~193~{\AA} but very weak. To look closely at
the appearance of transverse oscillations, five artificial straight
slits (S1$-$S5), which are almost perpendicular to their loop axes,
are chosen to generate the time-distance (TD) maps in
Figures~\ref{slit1} and \ref{slit2}. The magenta arrows outline the
positions and directions of those five slits. Next, one artificial
straight slit (S6) that is perpendicular to the solar prominence is
chosen to investigate the prominence oscillation. In order to
investigate the propagation of the EUV wave, two artificial straight
slits (S7 and S8) are chosen to generate the TD maps in
Figure~\ref{slit4}. In order to reduce the noise, the straight slits
S1$-$S6 are averaged over a 3$^{\prime\prime}$ width, while the
slits S7 and S8 are averaged over a width of 30$^{\prime\prime}$.
Also, a jet-like structure with the dark feature is observed at
AIA~211~{\AA} and 193~{\AA}, as indicated by the green arrow. The
online animations of v\_465.mp4 and v\_aia.mp4 show the evolution of
coronal loops and prominence and the associated EUV wave during
$\sim$17:48$-$18:15~UT. The animations are separated because that
the time resolutions of SDO/AIA and SUTRI are different.

\subsubsection{Loop oscillations}
\begin{figure}
\centering
\includegraphics[width=0.8\linewidth,clip=]{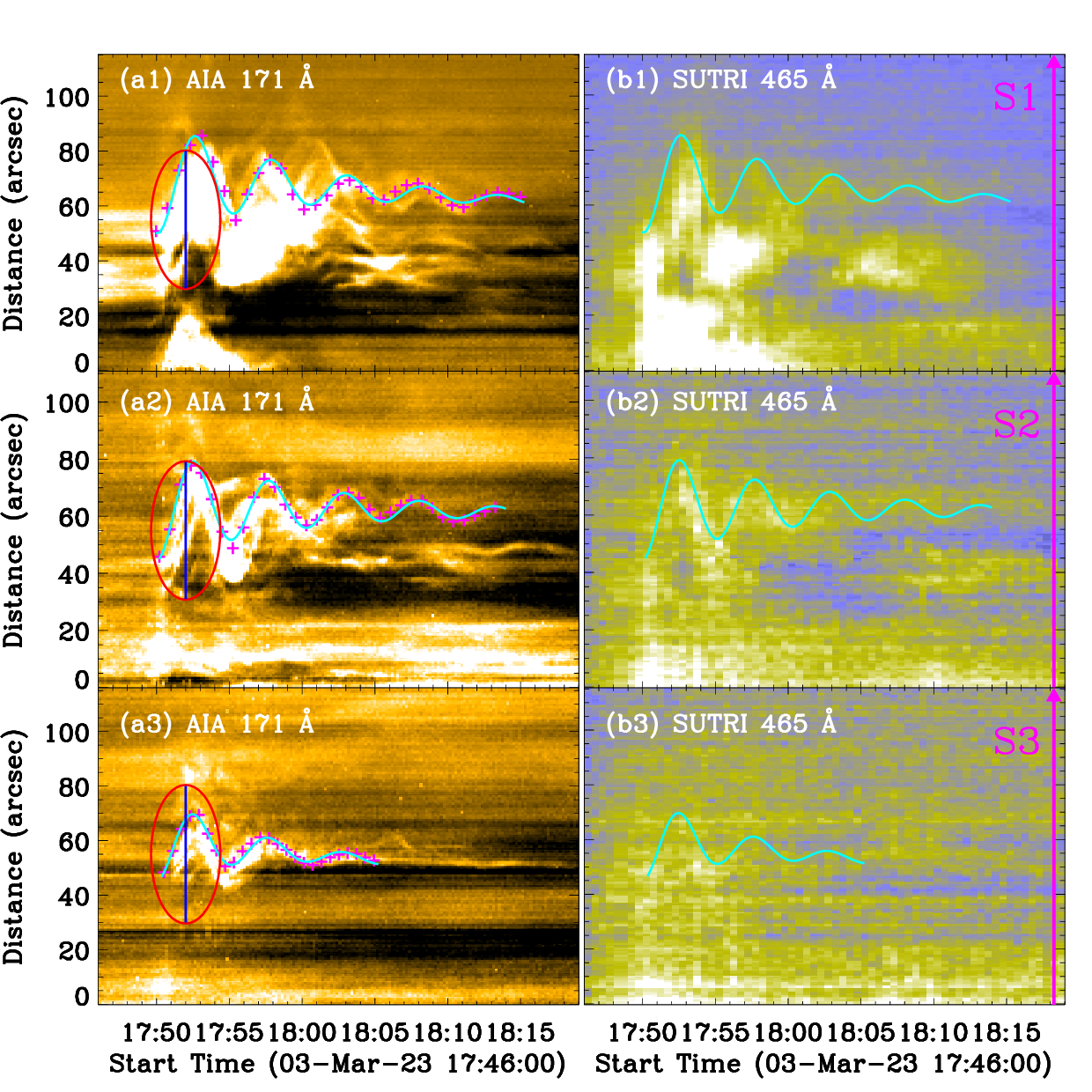}
\caption{TD maps showing the transverse oscillation of the short
loop (L1) at three slits (S1$-$S3), as observed in passbands of
AIA~171~{\AA} (a1$-$a3) and SUTRI~465~{\AA} (b1$-$b3). The magenta
pluses (`+') indicate the edges of the oscillating loop, whereas the
cyan curves represent the best-fitting results. The blue line inside
the red ellipse marks the peak time in the GOES SXR light curve. The
magenta arrow outlines the slit direction. \label{slit1}}
\end{figure}

In this study, only the short loop (L1) has a complete loop profile,
namely, the loop apex and two footpoints can be seen in passbands of
AIA~171~{\AA} and SUTRI~465~{\AA}. Thus, three straight slits
(S1$-$S3) that are nearly perpendicular to the loop axis are used to
generate TD maps, as shown in Figure~\ref{slit1}. Herein, the three
slits are selected at the locations that are close to the loop apex
(S2) and two footpoints (S1 and S3), respectively. Moreover, they
are also selected at the positions where there are less overlaps
with neighboring loops. In those TD maps, one can immediately notice
a transverse oscillation with apparent decay. For a typical
transverse oscillation, the oscillating positions are usually
identified as the loop centers with the Gaussian fitting method
\cite{Mandal21,Zhong22}. But it is difficult to use this method when
some overlapping loops simultaneously exist in the TD map
\cite{Anfinogentov15,Goddard16,Gao22}. Herein, the oscillation
positions are manually identified as the edge profiles of the
transverse oscillation in the passband of AIA~171~{\AA}
\cite{Gao22,Li23a}, as indicated by the magenta pluses (`+') in
panels~(a1)$-$(a3). Those transverse oscillations seem to decay
rapidly, and thus, a sine function combined with a decaying term and
a linear trend is applied to fit the observed transverse oscillation
\cite{Nakariakov99,Su18,Li23a}, as shown by Eq.~\ref{eq1}:

\begin{equation}
  A(t)=A_m \cdot \sin(\frac{2 \pi}{P}~(t-t_0)+ \phi ) \cdot e^{-\frac{(t-t_0)}{\tau}}+ b \cdot (t-t_0) + C,
\label{eq1}
\end{equation}
\noindent Here $A_m$ denotes to an initial displacement amplitude,
$P$ and $\tau$ represent the oscillation period and decaying time,
$t_0$ , $\phi$ and $C$ refer to the onset time, initial phase and
location of the decaying oscillation, while $b$ is a constant that
stands for the drifting speed of the oscillating loop in the
plane-of-sky. Then, the velocity amplitude ($v_m$) can be determined
from the derivative of the displacement amplitude
\cite{Li22a,Li23,Petrova23}, namely, $v_m=2\pi \cdot \frac{A_m}{P}$.
In Figure~\ref{slit1}, the over-plotted cyan curve in each panel
represents the best-fitting result using Equation~\ref{eq1}, which
appears to match well with the oscillating skeleton of the coronal
loop. Notice that the cyan curves in right panels are exactly the
same as those in left panels. Table~\ref{loop} (i.e., columns~2$-$4)
shows the main fitting parameters in the transverse oscillation of
the short loop at three slits.

\begin{table}
\caption{Fitting parameters of oscillating loops and prominence.}
\label{loop}  \centering   \tabcolsep 10pt
\begin{tabular}{c c c c c c c}
\toprule[1.1pt]
Slit                    &  S1     & S2       &  S3   &  S4     &   S5     &  S6      \\
$P$ (minutes)           &  5.2    &  5.2     &  5.2  &  13.5   &  29.3    &  18.5       \\
$\tau$ (minutes)        &  9.2    &  9.9     &  6.1  &  30.7   &  23.9    &  73.6     \\
$A_{\rm m}$ (Mm)        &  16.8   &  14.7    &  11.1 &  15.2   &  35.4    &  6.5     \\
$v_{\rm m}$ (km~s$^{-1}$) &  338  &  296     &  223  &  118    &  127     &  37    \\
$\frac{\tau}{P}$        &  1.8    &  1.9     &  1.2  &  2.3    &  0.8     &  4.0     \\
\bottomrule[1.1pt]
\end{tabular}
\end{table}

The loop L1 shows visible transverse oscillation, as shown in
Figure~\ref{slit1} and online animations of v\_465.mp4 and
v\_aia.mp4. By seeing the associated animation, one can see that the
loop L1 has a sheared shape, which is very close to the eruption
center. It is pushed away by the erupting structure before the
oscillation process. That is, a pushing process of the expanding
loop occurs during about 17:50~UT to 17:54 UT, which is closely
related to the loop oscillation. The transverse oscillation within
apparent decay is most visible in the passband of AIA~171~{\AA}, and
it becomes weak but still can be seen in the passband of
SUTRI~465~{\AA}, suggesting that the oscillating loop has a
multi-thermal structure. On the other hand, the decaying oscillation
of loop L1 at three different slits starts at nearly the same time,
and it is accompanied by the eruption of the X2.1 flare. This
observational fact implies that the loop oscillates in-phase along
the loop length, and thus the decaying oscillation could be in the
axial fundamental mode of a standing wave. The initial oscillation
amplitude also decreases when the cut slit is far away from the
flare region.

\begin{figure}
\centering
\includegraphics[width=0.8\linewidth,clip=]{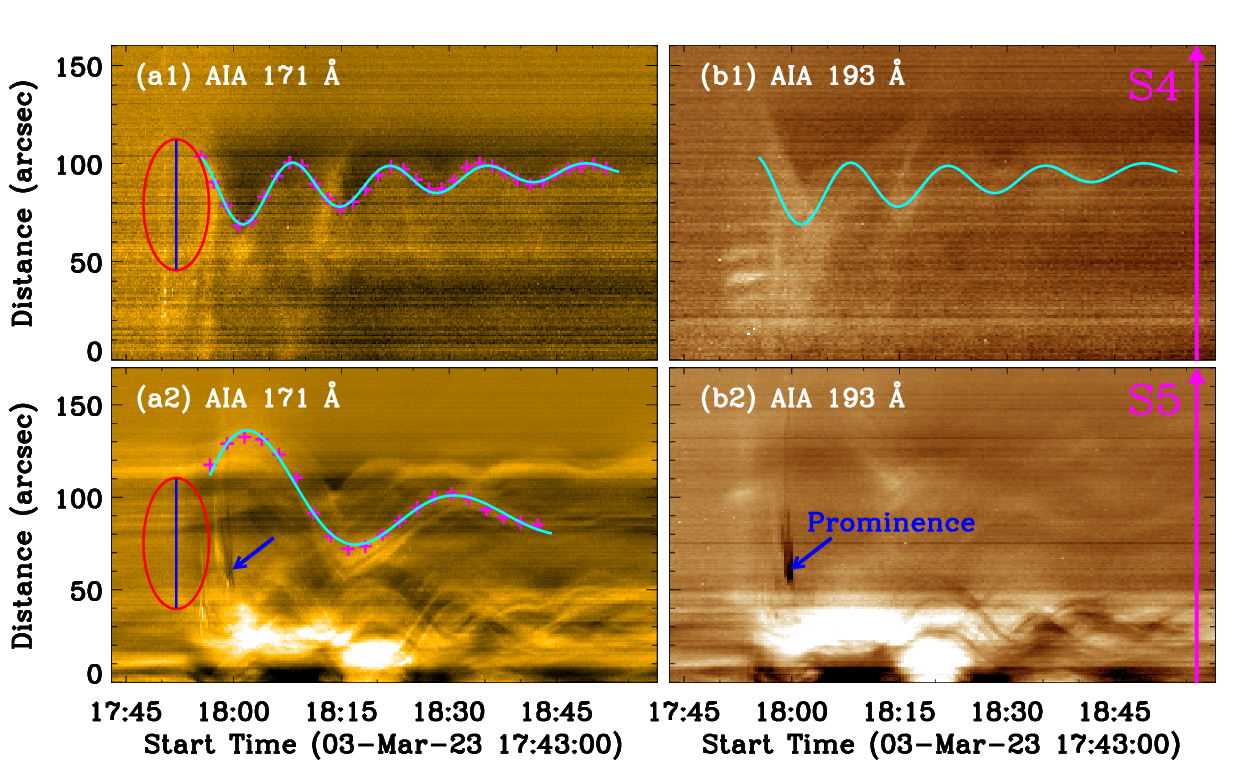}
\caption{Similar to Figure~\ref{slit1}, but showing the transverse
oscillations of two long loops (L2 and L3) at slits S4 and S5,
respectively. The blue line inside the red ellipse marks the peak
time in the GOES SXR light curve. The blue arrow indicates the
prominence. \label{slit2}}
\end{figure}

The loops L2 and L3 are very diffuse, both of which shows
unidirectional motions before the oscillation process, namely, the
loop L2 displays a northward motion, and the loop L3 reveals a
southward motion. These two loops are visible at AIA~171~{\AA},
become weak at AIA~193~{\AA}, and disappear at AIA~211~{\AA} and
SUTRI~465~{\AA}, as shown in Figure~\ref{img}. Therefore,
Figure~\ref{slit2} plots TD maps for loops L2 and L3 in passbands of
AIA~171~{\AA} and 193~{\AA}. For each loop, only one straight slit
is chosen. Similar to the loop L1, the two diffuse loops exhibit
transverse oscillations within apparent decay at AIA~171~{\AA}. So
we manually identify the edge profiles as the oscillating positions
and fit them with Eq.~\ref{eq1}, as shown by the magenta pluses and
cyan curves. The fitting parameters are shown in Figure~\ref{loop},
and they match well with each other, confirming that they are
decaying oscillations. In the passband of AIA~193~{\AA}, the
decaying oscillation of loop L2 becomes very weak but can be
identified, but it is hard to see the decaying oscillation of loop
L3, as shown in the right panels of Figure~\ref{slit2}. The decaying
oscillations of loops L2 and L3 both follow the X2.1 flare. After
all, the flare SXR emissions attain their peaks earlier than the
onset of the oscillations in the examined loops, as indicated by the
blue line inside a red ellipse. A solar prominence appears in the TD
maps at AIA~171~{\AA} and 211~{\AA}, as indicated by the blue arrow.

\subsubsection{Prominence oscillations}
\begin{figure}
\centering
\includegraphics[width=0.8\linewidth,clip=]{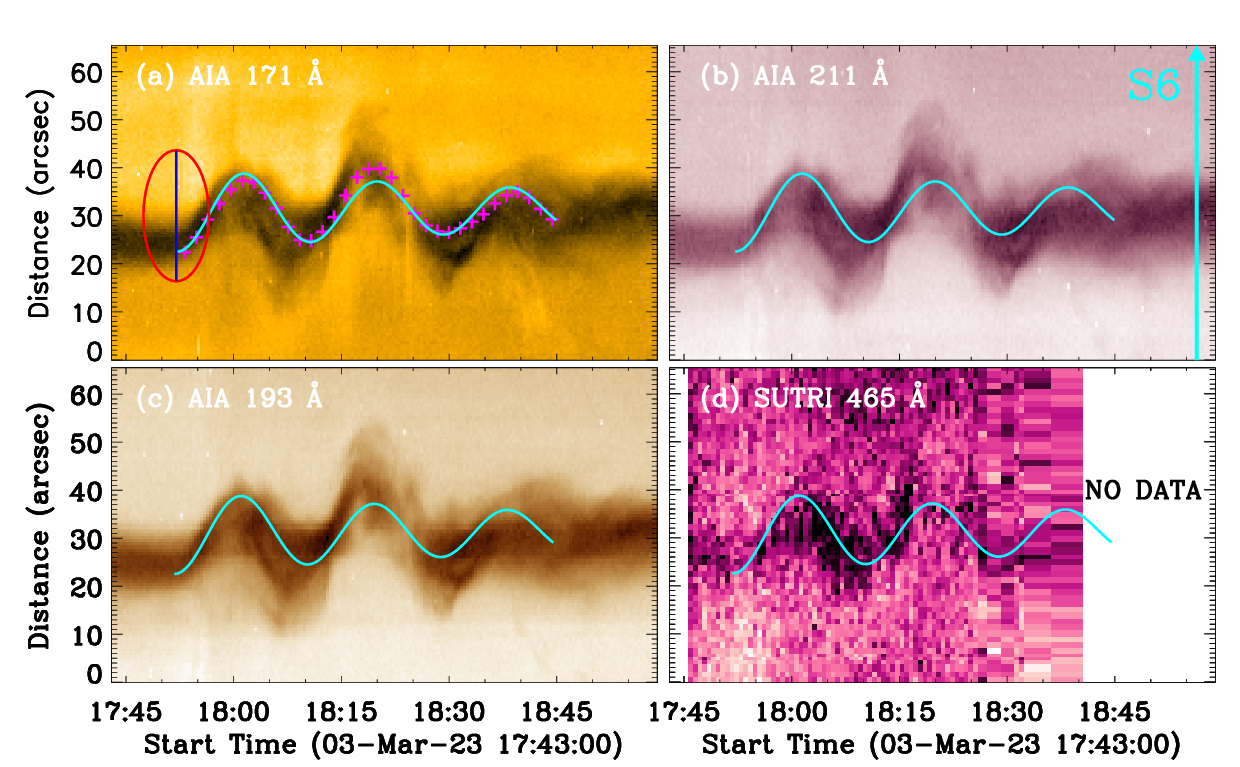}
\caption{TD maps showing the transverse oscillation of the
prominence at the slit S6, as observed in passbands of AIA~171~{\AA}
(a), 211~{\AA} (b), and 193~{\AA} (c), and SUTRI~465~{\AA} (d). The
magenta pluses (`+') indicate the edges of the oscillating loop,
whereas the cyan curve represent the best-fitting result. The blue
line inside the red ellipse marks the peak time in the GOES SXR
light curve. The cyan arrow outlines the slit direction.
\label{slit3}}
\end{figure}

A small prominence can be found in the solar north-west limb, which
also exhibits the transverse oscillation, as shown in
Figure~\ref{img} and online animations. The prominence lies in the
south of the active region. It shows a southward motion before the
oscillation process, and begins to oscillate after the passing of
the expanding CME bubble. Figure~\ref{slit3} shows the TD maps at
the slit S6 that is perpendicular to the prominence in passbands of
AIA~171~{\AA}, 211~{\AA}, 193~{\AA}, and SUTRI~465~{\AA}. It should
be pointed out that the origin maps rather than the base-diffidence
maps are used to generate those TD maps, and there is not data after
$\sim$18:40~UT at SUTRI~465~{\AA}. Those TD maps appear to reveal a
transverse oscillation within at least three peaks, and they decay
slowly. Similar to the oscillating loops, the oscillation positions
are determined by the edge profiles of the prominence oscillation in
the passband of AIA~171~{\AA}, as indicated by the magenta pluses
(`+') in panel~(a). Then, the observed transverse oscillation of
prominence is fitted by using Eq.~\ref{eq1}, and it seems to match
well with the oscillating profile of the prominence, as indicated by
the cyan curve in panel~(a). Similarly, the decaying oscillation of
the prominence is accompanied by the eruption of the X2.1 flare.
Then, the fitting function (indicated by the cyan curve) is directly
overplotted in TD maps at AIA~211~{\AA} and 193~{\AA} and
SUTRI~465~{\AA}, and it seems to agree with the prominence
oscillation, as can be seen in panels~(b)$-$(d).

\subsection{EUV wave}
The online animations of v\_465.mp4 and v\_aia.mp4 show that the
X2.1 flare generates an expanding loop structure, and then excites a
coronal wave-like structure propagating on the solar surface, which
could be regarded as an EUV wave \cite{Liu14,Warmuth15}. From the
online animations, we find a bright but weak wavefront  after the
expanding loop structure in passbands of AIA~211~{\AA}, 171~{\AA},
and SUTRI~465~{\AA}, then it travels fast with the arc-shaped
profile. A dark dimming region appears behind the wavefront, which
could be attributed to the significant reduction of coronal
densities. At the same time, a jet-like structure (indicated by the
green arrow in Figure~\ref{img}) is visible behind the wavefront,
which might be regarded as an erupting flux rope, as can be seen in
the animations.

\begin{figure}
\centering
\includegraphics[width=0.8\linewidth,clip=]{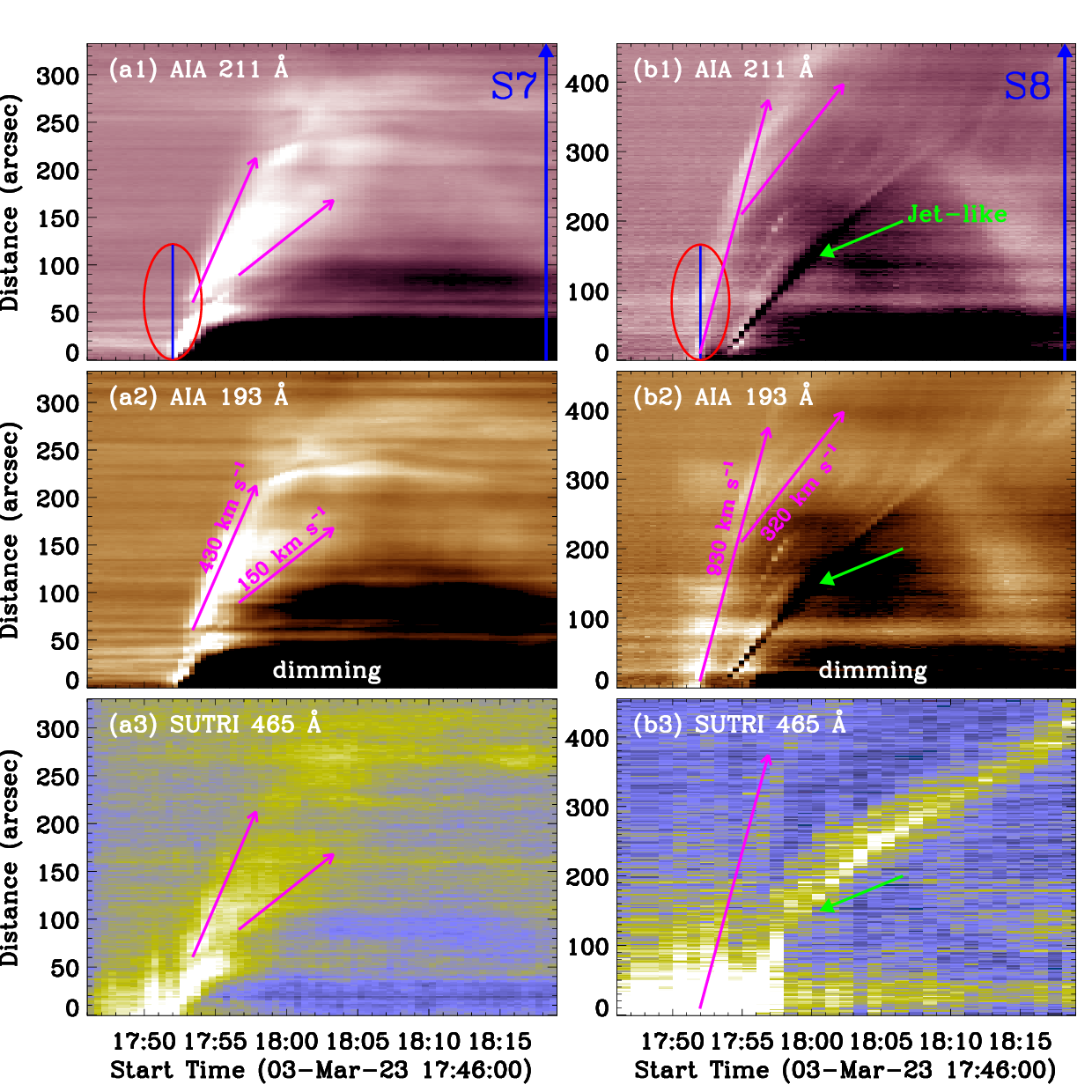}
\caption{TD maps showing the EUV wave at slits S7 and S8, as
observed at AIA~211~{\AA}, and 193~{\AA}, and SUTRI~465~{\AA}. The
magenta arrows show the EUV wave, and the green arrow indicates the
jet-like structure. The blue line inside the red ellipse marks the
peak time in the GOES SXR light curve. The blue arrow outlines the
slit direction. \label{slit4}}
\end{figure}

To illustrate the evolution of the EUV wave, Figure~\ref{slit4}
presents the TD maps at two slits (S7 and S8) that are along the EUV
wave propagation. Herein, the two slits are averaged over a
30$^{\prime\prime}$ width, so that the bulk of this EUV wave can be
covered as much as possibly. Figure~\ref{slit4}~(a1)$-$(a3) show the
TD maps along the slit S7 that is mainly inside the solar limb at
AIA~211~{\AA}, 193~{\AA}, and SUTRI~465~{\AA}, and they all reveal
the bright inclined features, which could be the indicator of the
EUV wave propagation. It is interesting that the EUV wave shows
obviously two components, namely, a slow component followed by a
fast one, as marked by the two magenta arrows. Then, the EUV wave
velocities could be determined by the slopes of the magenta lines,
which are equal to $\sim$430~km~s$^{-1}$ and $\sim$150~km~s$^{-1}$,
respectively. The velocity of the fast component is roughly three
times larger than that of the slow one. Similarly, the EUV wave
within two components is also detected in the TD maps along the slit
S8 that is outside the solar limb, as shown in
Figure~\ref{slit4}~(b1)$-$(b3). Their velocities are measured to be
about $\sim$930~km~s$^{-1}$ and $\sim$320~km~s$^{-1}$, which also
have a three-times difference. Here, a jet-like structure as marked
by the green arrow is simultaneously seen in passbands of
AIA~211~{\AA}, 193~{\AA}, and SUTRI~465~{\AA}. The jet-like
structure exceeds the FOV of our observational data, as shown in
Figure~\ref{slit4}~(b3), indicating that it could escape from the
solar surface and propagate into the interplanetary space. We also
notice that the EUV wave is accompanied by the eruption of the X2.1
flare. Also, a dark coronal dimming region appears after the
eruption of the EUV wave.

\section{Discussions}
Transverse oscillations are frequently observed in coronal loops,
flare loops, prominences or filaments, and even above the sunspot
\cite{Asai12,Goddard16,Li20,Nechaeva19,Zhang20z,Zhong23,Yuan23}. In
this study, we investigate the transverse oscillations in three
coronal loops, which are all associated with an X2.1 flare. They
could last for several wave periods and reveal a significant decay
in the displacement amplitude. The initial displacement amplitudes
are measured to be in the range of $\sim$11.1$-$35.4~Mm, and they
are found to decay rapidly. This is consistent with previous results
for the decaying oscillation in coronal loops
\cite{Aschwanden99,Nakariakov99,Li17,Su18}. The oscillation periods
are estimated to be from about 5.2~minutes to 29.3~minutes. Such
large differences of periods could be attributed to the various
lengths of oscillating loops, since the oscillation periods of the
standing kink wave strongly depend on their loop lengths
\cite{Anfinogentov15,Goddard16,Li23}. The same period of 5.2~minutes
is measured in the oscillating loop L1 at three different positions,
and there is not apparent phase difference at those three slits,
confirming that the observed transverse oscillation is indeed the
fundamental mode of a standing wave. The ratio ($\frac{\tau}{P}$)
between the decaying time and oscillation period is measured to be
about 0.8$-$2.3, which is similar to previous results for the
decaying kink oscillations, such as an average ratio of about 1.79
\cite{Nechaeva19}. All our observations suggest that the observed
transverse oscillations of coronal loops are basically standing kink
waves. This is different from our pervious observation, namely, the
traveling kink pulse of coronal loop is initially triggered by a
solar flare, and then it evolved to a standing kink wave
\cite{Li23a}. In this study, we do not find such evolution, although
both of them appear after the eruption of major flares. At the same
time, the transverse oscillation is also detected in a solar
prominence, and it decays slowly. The ratio between the decaying
time and the oscillation period is measured to be $\sim$4.0, which
is close to those detected in coronal loops. Thus, it could also be
interpreted as the decaying kink wave.

We also study an EUV wave, which could be excited by the expanding
loop structure. The EUV wave can be simultaneously seen in passbands
of AIA~211~{\AA}, 193~{\AA}, and SUTRI~465~{\AA}, which agrees with
previous observations \cite{Hou23}. We do not show the EUV wave at
AIA~171~{\AA}, because it is obscure, which might be attributed to
the multiple coronal loops at AIA~171~{\AA}, i.e., the loop L3. In
agreement with previous numerical and observational results
\cite{Chen02,Chen11,Liu14,Shen14a}, two components of the EUV wave
are simultaneously observed, that is, the fast and slow components.
The fast component is interpreted as the fast-mode MHD wave or the
shock wave. The slow component wave-like signature is a non-wave
component, and it may be induced by the reconfiguration of magnetic
field lines that caused by the associated CME. It is just a visual
effect caused by the disturbance resulted from the stretching
magnetic field lines. A halo
\href{https://www.sidc.be/cactus/catalog/LASCO/2\_5\_0/qkl/2023/03/CME0015/CME.html}{CME}
was detected to start at 18:24~UT, which could be associated with
the EUV wave. The non-wave component always appears later than the
fast-mode wave component. A jet-like structure appears and
propagates after the EUV wave, which could be the erupting flux rope
that acts as the core of the associated CME
\cite{Shen19,Shen19b,Shen21}.

\begin{figure}
\centering
\includegraphics[width=0.8\linewidth,clip=]{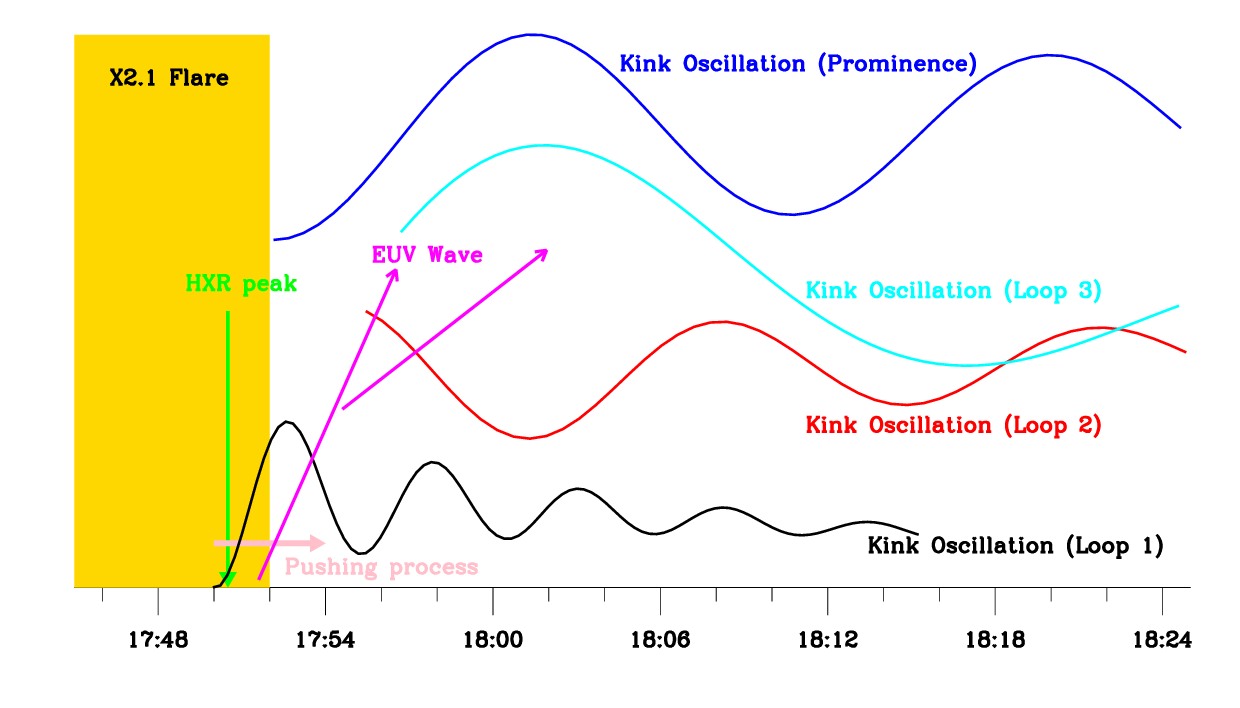}
\caption{Sketch of the timeline to show a plausible scenario of kink
oscillations of three coronal loops (L1$-$L3) and one prominence,
and an EUV wave, which are associated with an X2.1 flare
(gold-shaded box). The green arrow marks the HXR peak time, and the
pink arrow outline the pushing process of the expanding loop.
\label{timl}}
\end{figure}

Both the EUV wave and decaying kink oscillations are associated with
an X2.1 flare, but what is their relationship? To illustrate this
issue, Figure~\ref{timl} draws the time line of the studied events.
The gold-shaded box represents that the period lasting from 17:45~UT
to the SXR peak time of the X2.1 flare, and only the beginning parts
of some kink oscillations (i.e., loops L2 and L3, and prominence)
are shown, since we mainly focus on their triggers. It can be seen
that the kink oscillation of the short loop L1 (black curve) first
appears after the X2.1 flare eruption, which is reasonable because
that the short loop L1 is closest to the flare region. Moreover, the
beginning time of this kink oscillation is nearly coincident with
the peak time of the HXR pulse recorded by Insight-HXMT, as marked
by the green arrow. Our observation is consistent with previous
findings for the decaying oscillations of coronal loops
\cite{Zhang20z}, implying that the trigger of kink oscillations
could be associated with a strong energy release rather the very
beginning energy release via magnetic reconnection. As can be seen
from the online animation, the kink oscillation of the short loop is
very likely to be driven by the pushing of the expanding loop. On
the other hand, the start time of kink oscillations of loops L2 and
L3 (red and cyan curves) are obviously later than that of the loop
L1, mainly because that those two coronal loops are far away from
the flare region. This also implies that the kink wave is probably
traveling in various coronal loops \cite{Li23a,Lib23}. For the kink
oscillation of the long loop (L3) and the prominence, they are
possibly triggered by the disturbance of the CME, because that they
both start after the passing of the expanding CME bubble, and the
prominence is likely the part of the long loop connecting the
eruption center. The initiation of the kink oscillation of loop L2
is unclear, it might be due to the interaction of the EUV wave,
although the propagation of the EUV wave in this region is very
weak. However, we can find that the loop L2 shows a northward motion
firstly and then begins the oscillation process, indicating that the
disturbance comes from the south direction.

In Figure~\ref{timl}, we notice that the decaying kink oscillations
and the EUV wave are almost simultaneously accompanied by the X2.1
flare, especially for the decaying kink oscillation of the loop L1,
for which the excitation mechanism is unlikely to involve the EUV
wave as an intermediary. This is different from previous
observations, for instance, the EUV waves triggered by solar flares
often result in transverse oscillations of the remote coronal loops
and prominences or filaments when they propagate across those
loop-like structures \cite{Asai12,Xue14,Shen17,Devi22,Zhang22}.
However, we do not find such relationship between kink oscillations
and EUV waves, suggesting that their triggered sources are same,
i.e., the expanding loop structure associated with the X2.1 flare.

\section{Summary}
Using the imaging observations measured by SUTRI, SDO/AIA, and
CHASE, combined with the X-ray light curves recorded by GOES and
Insight-HXMT, we investigate flare-associated kink oscillations and
EUV waves, and the objective being to shed more light on the broad
subject as to how low-frequency waves are excited in the solar
atmosphere. Our main conclusions are summarized as follows:

(1) The decaying kink oscillations and the EUV wave are
simultaneously observed in passbands of AIA~171~{\AA}, 193~{\AA},
211~{\AA}, and SUTRI~465{\AA}, and they are all associated with an
X2.1 flare.

(2) The decaying kink oscillations are found in three coronal loops
and one solar prominence, and they can be interpreted as the
standing kink waves. The kink oscillation of the loop L1 is
identified as an axial fundamental kink mode, and it could be driven
by the pushing of the expanding loop. The kink oscillations of the
loop L3 and the prominence are possibly excited by the disturbance
of the CME, and the kink oscillation of the loop L2 might be due to
the interaction of the EUV wave.

(3) Two component EUV wave-like structures are simultaneously
observed in passbands of AIA~211~{\AA} and 193~{\AA}, and
SUTRI~465{\AA}. Their speed ratio is roughly equal to three. The
fast component could be a fast-mode MHD wave, which should be driven
by the lateral rapid expansion of the erupting loops. The slow
component is non-wave component, which might be just a visual effect
caused by the disturbance resulted from the stretching magnetic
field lines.

(4) A jet-like structure is observed to propagate after the EUV
wave, which could be regarded as the erupting flux rope that acts as
the core of the CME.

\section*{Acknowledgements}
We thank the two referees for their constructive suggestions and
detailed comments. This work was supported by the National Key R\&D
Program of China 2021YFA1600502 (2021YFA1600500), NSFC under grants
11973092, 12073081, 12003064, 12333009. D. Li is also supported by
the Surface Project of Jiangsu Province under the grant BK20211402
and Yunnan Key Laboratory of Solar Physics and Space Science
(202205AG070009) under the grant YNSPCC202207. SUTRI is a
collaborative project conducted by the National Astronomical
Observatories of CAS, Peking University, Tongji University, Xi'an
Institute of Optics and Precision Mechanics of CAS and the
Innovation Academy for Microsatellites of CAS. The CHASE mission is
supported by China National Space Administration (CNSA).

\vspace{5mm} 

\end{document}